\documentclass[times,10pt,twocolumn]{article}
\usepackage{latex8}
\usepackage{times}

\usepackage{url}
\usepackage{graphicx}
\usepackage{moreverb}

\newcommand{\myTable}[1]{Table~\ref{#1}}

\newcommand{\myhead}[1]{\par\noindent\textbf{#1}~~}

\newcommand{\mySection}[1]{\Section{#1}}

\begin{document}

\title{
  Policy and Legal Challenges of Virtual Worlds and Social Network Sites
}

\author{
  Holger M. Kienle\\
  University of Victoria\\
  Victoria, Canada\\
  {\tt hkienle@acm.org}
  \and
  Andreas Lober\\
  RAe Schulte Riesenkampff\\
  Frankfurt am Main, Germany\\
  {\tt alober@schulte-lawyers.de}
  \and
  Hausi A. M\"uller\\
  University of Victoria\\
  Victoria, Canada\\
  {\tt hausi@cs.uvic.ca}
}

\date{}
\maketitle
\thispagestyle{empty}

\begin{abstract}
  This paper addresses policy challenges of complex virtual
  environments such as virtual worlds, social network sites, and
  massive multiplayer online games. The complexity of these
  environments---apparent by the rich user interactions and
  sophisticated user-generated content that they offer---poses unique
  challenges for policy management and compliance. These challenges
  are also impacting the life cycle of the software system that
  implements the virtual environment.  The goal of this paper is to
  identify and sketch important legal and policy challenges of virtual
  environments and how they affect stakeholders (i.e., operators,
  users, and lawmakers).  Given the increasing significance of virtual
  environments, we expect that tackling these challenges will become
  increasingly important in the future.
\end{abstract}

\mySection{Introduction and Background}\label{sec:Intro}

In this paper, we explore the distinct characteristics of virtual
environments, and identify the legal and policy challenges that they
pose. We argue that the complexity of these environments and the
richness of interactions that they offer result also in an increase of
complexity in the management, compliance, and auditing of policy and
legal requirements.

In the following, we address complex computer-generated environments,
namely virtual worlds (VWs), social network sites (SNSs), and Massive
Multiplayer Online Games (MMOGs) \cite{BE:JCMC:07} \cite{Lober:07}.
Examples of VWs are Second Life, There, and Habbo Hotel;
examples of SNSs are Facebook, LinkedIn, and Xing; and examples of
MMOGs are World of Warcraft, MapleStory, and RuneScape.
In the subsequent discussion we will use the term \emph{virtual
  environment} (VE) when discussing issues that apply to VWs, SNSs,
and MMOGs. VEs have in common that they enable multiple users to
interact and collaborate in a complex computer-generated environment.

VEs are increasingly gaining significance in terms of numbers of users
and generated revenue.\footnote{For example, Xing claims that every day
  5.7 million people use their platform. The most popular MMOG is
  World of Warcraft, whose owners claim to be generating 1 billion USD
  in revenue per year with over 10 million subscribers
  \cite{DL:SCCHTLJ:08}.
} As a result, policy and legal issues are becoming more and more
important for the stakeholders of VEs (i.e., users/players,
providers/operators, and lawmakers/regulators).

VEs are diverse in the sense that they (1) attract people based on a
wide range of different interests such as shared hobbies, sports,
religion, and sexual interests, (2) have different purposes such as
game-playing, socializing or business, (3) support different
interaction patterns such as real-time 3D interactions or asynchronous
communication based on message boards, and so on
\cite{BVFSD:IJIDM:08}.  Consequently, there is no crisp definition of
a VE that allows one to draw a clear boundary. In fact, one may view a
simple listserv as a social network and, as such, as a VE
\cite{PJ:FirstMonday:06}.
A common characteristic of VEs is that there is an emerging culture
shaped by social interactions of its members in a virtual environment.

In the following, we contrast VEs with different kinds of web sites.
The architecture of the World Wide Web has many characteristics that
are similar to VEs and as a result many VEs are based on the Web's
infrastructure. For example, many social networks are implemented as
web sites, and some 3D worlds run with web browser plug-ins (e.g.,
Habbo Hotel runs in Adobe's Shockwave player).
For discussion purposes, we introduce a classification of web sites,
which groups the sites with roughly increasing sophistication in terms
of content and interaction models:
\begin{description}
\item[brochure-ware:] These sites provide information that users can
  browse (e.g., to obtain information about products and services that
  they can obtain off-line) \cite{TH:ICSE:01}. Users do not have to log
  on to the site and the site is static in the sense that it looks the
  same for all users.
\item[e-commerce:] These sites are run by companies that sell products
  online. They may be pure online retailers (\emph{e-tailers}) or have
  a \emph{clicks-and-bricks} hybrid business model \cite{PG:CACM:03}.
  To place orders, users have to create an account.
\item[Web 2.0:] These sites are characterized by sophisticated
  functionality that often rival shrink-wrapped software products.
  These sites typically offer a participatory and interactive user
  experience \cite{CK:FirstMonday:08}. Importantly, these sites have
  user-generated content where users are \emph{conducers}, that is,
  they ``both consume creative works and simultaneously add creative
  content to those same works'' \cite{Reuveni:SSRN:08}.
\end{description}
The above classification is an idealization because concrete web sites
typically have features that blur into other groups. For example, a
brochure-ware site may have a form or questionnaire that users can fill
out to provide feedback to the site operator, and e-commerce sites
often have some kind of personalization (e.g., Amazon's wishlists) or
user-generated content (e.g., book reviews of users).

\mySection{Policy and Legal Challenges}\label{sec:}

\myhead{Stakeholders:}
The two most important stakeholders of VEs are its operators and its
users. The relationships between both stakeholders are primarily
governed by the policies embodied in the terms of use statement and
the privacy policy.
Policies often interact with legal requirements. In this context,
there are additional stakeholders such as lawmakers that create
regulations, and courts that create case law. Hence, policy challenges
have to factor in legal requirements as well (e.g., privacy policies
are constrained by privacy regulations).
There are also organizations such as the Virtual Policy Network
(\url{virtualpolicy.net}) that aim at bringing together
stakeholders from government, academia, and industry.
In the following, we discuss a number of selected issues that concern
the interaction of operators and users as well as lawmakers. These
issues are meant to expose challenges that are particularly relevant
to VEs.

\myhead{Legal considerations:}
In the early days of the Web, it was often perceived as being free and
unregulated \cite{Lessig:99}. This perception has gradually changed
with increasing maturity and commercialization of the Web. VEs have
made a similar development in this respect.
Many legal issues of the Web and of VEs are addressed by existing laws
and case law. However, there are also specific acts (e.g., U.S.'s
COPPA) and policies (e.g., ICANN's UDRP) that have been enacted for
cyberspace. It remains to be seen if lawmakers will become active for
VEs. On April 1st, 2008 a first hearing by the \emph{Subcommittee on
  Telecommunications and the Internet} was held on policy concerns of
VEs.\footnote{\url{http://energycommerce.house.gov/cmte_mtgs/110-ti-hrg.040108.VirtualWorlds.shtml}}

Almost all legal issues that exist in real life are potentially
applicable to VEs; this holds especially for 3D VWs. The only question
is how to map virtual incidents to applicable law: Killing a human is
not the same as killing an avatar, so the latter is not being
considered murder (even though there may be other repercussion of such
an act depending on the VW), smoking pot in a VW is not a use of
illegal drugs (but may be considered promoting drug abuse), and sexual
acts with kid-faced avatars is not child abuse (but potentially child
pornography).
Prominent legal issues that arise in all kinds of web sites and VEs
are copyright and trademark, especially if they allow user-generated
content \cite{DL:SCCHTLJ:08}. For e-commerce sites and VEs there is
also taxation, fraud and money laundering. In VEs these issues surface
if the world has an economic model involving virtual money and users
that can own virtual property \cite{LH:UPLS:03} \cite{LH:NYSLR:04}.
Virtual money (e.g., Second Life's Linden Dollars, There's Therebucks,
and Entropia Universe's PED) is real in the sense that they can be
exchanged for real money and vice-versa.
If the VW allows (real-time) user interactions (e.g., avatar movements
in 3D and voice chat) there is also the possibility of harassment,
assault, and libel.
An overarching legal issue is jurisdiction because many sites and VEs
are not constrained by national boundaries. For instance, VEs are
often implemented as server farms that are located throughout the
world. As a consequence, the access, storage, and replication of data
may be constrained by different data protection laws.
If the VE has a virtual  currency and enables gambling, there may be
complex legal questions depending on the locations of the operator,
its servers, and the users.
Interestingly, operators can try to segregate or exclude users. Second
Life has a dedicated Teen Area where users are required to be between
13--17 years of age. E-commerce sites can exclude users via
restricting shipping to postal addresses in certain countries.

\myhead{Complexity:}
From the users' perspective, policies are important because they spell
out their rights and obligations. Unfortunately, these policies are
often difficult to read and understand (e.g., privacy polices in the
healthcare domain \cite{AEVJG:IEEESP:07}).
Furthermore, VEs offer rich user interactions and business models that
have to be reflected in their policies. As a result, such policies are
comparably complex.
While brochure-ware sites can be satisfied by covering only general
issues (e.g., license to use, disclaimer, linking, and intellectual
property), e-commerce sites also have to address issues such as order
acceptance, pricing information, exporting of goods, and disclaimers
for special goods such as medicines. Similarly, VEs have to cover
issues that are unique to their environment; for example, Second
Life's terms of use addresses trading of its virtual currency.

\begin{table}[tb]
  \begin{tabular}[htb]{|l||l|r|r|}
  \hline
  Operator       & Type          & \# words & Flesch \\
  \hline
  \hline
  ge.com         & brochure-ware & 1576     & 56.3 \\
  wal-mart.com   & e-commerce    & 5056     & 57.4 \\
  secondlife.com & 3D world      & 7492     & 42.2 \\
  \hline
\end{tabular}
\centering
\caption{Examples of terms of use statements and their number of words and Flesch readability score}
\label{fig:ToU}
\vspace{-5mm}
\end{table}

\myTable{fig:ToU} shows statistics of three different terms of use
statements: a brochure-ware site (General Electric), an e-commerce
site (Wal-Mart), and a VW (Second Life). In these examples, more site
complexity translates into an increase in the size of the terms of
use. The Flesch readability\footnote{The Flesch Reading Ease measures
  how easy it is to read a text with a score from 0 to 100, where a
  lower score indicates a more difficult text. Scores of 50--59 are
  considered fairly difficult and 30--49 difficult.} for Second Life
indicates that the terms of use are significantly more difficult to
understand than for the other two sites.

Even though these policy statements are already complex, they cannot
hope to be all-encompassing. As a result, they represent an element of
uncertainty to both operators and users. It is an open question how to
minimize uncertainty in policies. The complexity of VEs may prompt
operators to look for novel approaches on how to represent and enforce
policies, and how to negotiate and contract policies with users.

\myhead{Compliance:}
Operators need to manage and enforce the policies, a fact which
represents a significant challenge in VEs. First, elements of policies
(expressed in natural language) have to be expressed as constraints in
the VW, which is ultimately realized in its code. However, mapping the
policies down to code and keeping both consistent in case one or the
other evolves is difficult to manage.
There are many examples of privacy violations caused by wrong
implementations of privacy features. For example, in Facebook
supposedly private annotations were made visible to all users
\cite{Goodin:Register:07}. In contrast, privacy of e-commerce sites is
comparably easy to express because no user is allowed to see any data
or interactions of other users.
Second, enforcement of policies is difficult in VEs because of the
high degree of freedom that users have in interacting with the
environment. For example, enforcement of intellectual property (IP) rights in a brochure-ware
site is relatively easy because the content publishers can be managed.
If simple user-generated content such as book reviews are allowed, the
IP violations can be limited by the form of expression (e.g., text
only, limited number of words). Furthermore, content such as text is
amenable to automated processing, and the content of web sites can be
crawled to look for policy violations. In contrast, ``crawling'' and
automated processing of the content of a 3D environment to ensure
compliance with policies is a much bigger challenge.

\myhead{Negotiation and balance:}
Another challenge of VEs is how to negotiate policies between
operators and users. Currently policies are drafted and put into
effect by operators without consulting users, and operators try to
reserve the right to change policies at will. This can result in
unbalanced policies that put users at a disadvantage. The following is
an excerpt of a legal notice from the web site of a large U.S.\@
corporation in 1998 (essentially
brochure-ware):\footnote{\url{http://web.archive.org/web/19980530081620/www.valero.com/html/legal_notice.htm}}
\vspace{-2mm}
\begin{quote}
  ``Any visitor to the Valero web site who provides information to
  Valero agrees that Valero has unlimited rights to such information
  as provided, and that Valero may use such information in any way
  Valero chooses. Such information as provided by the visitor shall
  be non-confidential.''
\end{quote}
In the past policies have been criticized if perceived as unbalanced.
For example, the first terms of use of Adobe's Photoshop Express
stated that users who uploaded pictures in effect
\vspace{-2mm}
\begin{quote}
  ``grant Adobe a worldwide, royalty-free, nonexclusive, perpetual,
  irrevocable, and fully sublicensable license to use, distribute,
  derive revenue or other remuneration from, reproduce, modify, adapt,
  publish, translate, publicly perform and publicly display such
  [pictures].''
\end{quote}
After this policy was widely criticized, Adobe made changes that
limited its rights to the pictures.
In contrast to most MMOGs, Second Life permits the creators of virtual
property to own their creations \cite{WIPOmag:07}. Second Life's terms
of use say explicitly:
\vspace{-2mm}
\begin{quote}
  ``You retain copyright and other intellectual property rights with
  respect to Content you create in Second Life, to the extent that you
  have such rights under applicable law.''
\end{quote}

Operators have to balance their desire to control and own
user-generated content and private data with the desire of users to
retain their own creations and to protect their privacy. However, when
users retain intellectual property of their creations, certain
challenges have to be faced when these creations become part of the
VE. For instance, if a user sells one of his virtual creations,
certain rights attached to it may have to be transferred or licensed
to the new owner;
and if users retain the copyright of their avatars, what about
screenshots with a commercial interest that are depicting them?
An unbalanced policy that is not freely bargained (i.e., a
\emph{contract of adhesion}) and that puts users at a clear
disadvantage increases the operator's risk that courts will find it
\emph{unconscientious}---and as a result may refuse to (partially)
enforce it \cite{Kunkel:MUEJL:02}.
Currently users have no negotiation power of policies (except via
lobbying and media coverage), even though operator-driven projects
such as BetterEULA (\url{bettereula.com}) provide a platform for user
input.  Also, European customer protection laws have been passed on
the assumption that end-consumers have no choice other than to accept
the policies imposed on them.
In the future, operators may want to offer personalized policies that
are semi-automatically negotiated.  Users and service providers could
state their privacy needs in machine-readable data for automated
negotiation of a privacy policy that is acceptable for both sides
\cite{MOL:LPPNaSDE:06}.
Again, this will result in increasing complexity for policy management
and compliance \cite{ABLY:CACM:07}. This complexity may be tackled
with policy-driven systems \cite{Barrett:POLICY:04}.

Generally, one can argue that users can switch VEs if they are not
happy with its policies, but there are significant barriers in
practice. For e-commerce sites there is a low cost to user to switch
operators (e.g., abandoning Barnes \& Noble in favor of Amazon)
because it requires to only open up a new account.  For social
networks, switching of sites (e.g., from Xing to LinkedIn) means
losing all the effort of populating ones profile and also ones social
identity. If users have heavily invested (also monetarily) in VEs
(e.g., purchase of land in Second Life, or building up an avatar in
World of Warcraft), switching is even more prohibitive (even though
Bartle points out that users could switch by selling and buying
avatars on eBay \cite[p.~111]{Lober:07}).
Finally, users tend to choose a VE for its content, not its policies.
There is no effective competition between operators of VEs for the
most user-friendly policy and as a consequence many stipulations that
are disadvantageous for users can be found in nearly all policies.

\myhead{Privacy:}
Privacy concerns are an important issue that serves as a good example
to expose policy challenges of VEs \cite{BE:JCMC:07}. On
brochure-ware sites there are only privacy issues of tracking the
movements of users on the site.  E-commerce sites have to protect
private data about users such as address and billing information. In
contrast, users create and expose all kinds of private information on
VEs, and VEs are also generating private data about users (via
profiling and mining techniques \cite{Hildebrandt:DuD:06}). Examples
of private information are user details (e.g., age, location, gender,
and testimonials), connectivity (e.g., friends and groups), content
(e.g., photos, commenting, and tagging) \cite{CK:FirstMonday:08}.
Facebook, for instance, supports the creation of all of the
aforementioned information.
Importantly, Quirchmayr and Wills make the point that ``the more data
we collect about a person, the more sensitive this data becomes,
because the increasing amount of available data allow to construct an
increasingly complete profile'' \cite{QW:TrustBus:07}. A less-welcome
scenario is that automated reasoning may create wrong knowledge about
a person, which is then difficult to purge or change
\cite[p.~152ff]{Taipale:YJOLT:04}.  Cormode and Krishnamurthy have
studied the unique characteristics of the Web 2.0 and conclude that
``there are significant challenges in allowing users to understand
privacy implications and to easily express usage policies for their
personal data'' \cite{CK:FirstMonday:08}.
VEs may push user monitoring and profiling to new levels. Even for 3D
worlds it seems feasible to record fine-grained movements and
interactions of avatars. Privacy concerns in VEs are similar to the
ones in real life. If the location data and history of a cell phone is
considered private, the same could be argued for an avatar---but there
may be sensible reasons for doing this (e.g., tracking of virtual
commerce transactions). It is currently difficult to assess for users
whether a VW's privacy policy and preference settings are adequate for
their personal perception of privacy.

\myhead{Evolution:}
Another challenge is the evolution of policies. As mentioned before,
an operator has a strong interest not to be restricted in any form
when making changes to polices of the VE as well as making changes to
the VE itself. To preserve consistency, a certain change in the VE may
mandate a corresponding change to its policy, and vice-versa.
In a sense, operators are the Gods of VEs because they have the means
to change its behavior as they see fit---in this respect, ``code is
law'' \cite{Lessig:99}. Indeed, Bartle, one of the pioneers of MMOGs,
argues that operators should be allowed to make drastic changes to a
VW, including its destruction, because users always have the option to
abandon it \cite[p.~114f]{Lober:07}). The risks that users of VEs face
have the following analogy: ``In the real world, those who make
investments in a country expose themselves to uniquely `sovereign'
risks because of the danger that the government might alter the laws
under which they claim to hold assets'' \cite{Grimmelmann:NYLSLR:04}.
However, whereas in real life the investors will probably not be in a
position to sue the sovereign, users of a VE can certainly sue its
operator.
The more users have invested in a VE and have come to depend on
certain behaviors of the VE, the more likely that they will sue if
they believe that a change in behavior constitutes a misconduct on the
side of the operator. In this respect, code is not the supreme law
because its evolution is constrained by policy.
For instance, there are users that derive significant revenue from
Second Life so that their ``business activities have been successful
enough to replace their real-life income,'' \cite{WIPOmag:07} as
exemplified by a user who claims to have earned \$1 million USD with
virtual property dealings.
If virtual property is in fact real as argued by Lastowka and Hunter
\cite{LH:UPLS:03}, actions by the operator that destroy or de-value
property may be actionable under law. Interestingly, Second Life is
indeed influencing its virtual real estate market by controlling the
supply rate of new
land.\footnote{\url{http://secondlife.reuters.com/stories/2008/06/19/linden-freezes-land-supply-as-prices-plummet/}}
This poses the question of the legal consequences if actions taken by
the operator---intentionally or unintentionally---cause a significant
de-valuation of all or some property. Lastly, this leads to the
question how operators would be able to terminate a highly developed
VW. Presumably, the operator would not have enough assets to cash-out
all users.
However, so far no VW that models a complex economy such as Second
Life has shut down.

\mySection{Conclusions}\label{sec:Concl}
\vspace{-1mm}

In this paper we have identified key challenges of virtual
environments with respect to the management, compliance, negotiation,
and evolution of policies.
We have contrasted challenges of virtual environments with the policy
issues faced by different groups of web sites (i.e., brochure-ware,
e-commerce, and Web 2.0), exposing that virtual environments exhibit
distinct characteristics that make policy issues particularly
challenging.

Important questions in this context are:
\begin{itemize}
\item How to ensure consistency among policies (e.g., polices embodied
  in terms of use statements and policies embodied in the code)? The
  complexity of virtual environments makes it difficult to keep
  policies consistent, and to define policies in code.
\item How to effectively enforce polices? On the one hand, there are
  technical challenges (e.g., automatically detecting a virtual
  trademark violation). On the other hand, the privacy of users has to
  be respected as well.
\item How to negotiate policies and how to give users more negotiation
  power? Given that policies express obligations of the user, a more
  balanced approach is needed so that unbalanced contracts of adhesion
  can be avoided.
\item How to evolve policies and the behavior of the virtual
  environment? In both cases, operators are constrained by user and
  legal considerations.
\item How to manage policies in an uncertain legal environment?
  Currently there is little case law to guide operators on how to meet
  legal requirements.
\end{itemize}

We believe that the increasing significance of virtual environments
and their unique characteristics deserve further exploration and
research of their policy issues by researchers in the legal,
governance, and computer science fields.

\bibliographystyle{latex8}
\bibliography{books,papers,local}

\noindent \includegraphics[scale=.6]{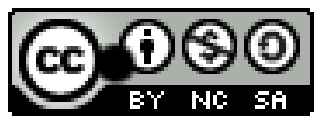}
\begin{minipage}[b]{.75\linewidth}
  {\tiny This work is licensed under a Creative Commons
    Attribution-Noncommercial-Share Alike 3.0 United States License.
    The license is available here:
    \url{http://creativecommons.org/licenses/by-nc-sa/3.0/us/}.}
  \baselineskip8pt
\end{minipage}

\end{document}